# A FIR-Survey of TNOs and Related Bodies

J. M. Bauer[1,2], P. F. Goldsmith[1], C. M. Bradford[1], A. J. Lovell[3]

[1]Jet Propulsion Laboratory, California Institute of Technology, Pasadena, CA, USA; [2] Infrared Processing and Analysis Center, California Institute of Technology, Pasadena, CA,USA; [3]Department of Physics and Astronomy, Agnes Scott College, Decatur, GA, USA

The small solar-system bodies that reside between 30 and 50 AU are referred to as the Trans Neptunian Objects, or TNOs. They comprise, in fact, the majority of small bodies within the solar system and are themselves a collection of dynamically variegated subpopulations, including Centaurs and Scatter-Disk Objects (SDOs), as well as “cold” (low-inclination and eccentricity) and “hot” (high eccentricity) classical Kuiper Belt populations (KBOs; Gladman et al. 2008). These minor planets are the reservoir of the comets that routinely visit our inner solar system, the short period comets, and so cloud the distinction between asteroids and comets. They are primordial material, unmodified by the evolution of the solar system and are the sources of volatile materials to the inner solar system (Barucci et al. 2008). Study of TNOs can thus inform us about the early history of the solar system, and how its composition has evolved over the time since it was formed.

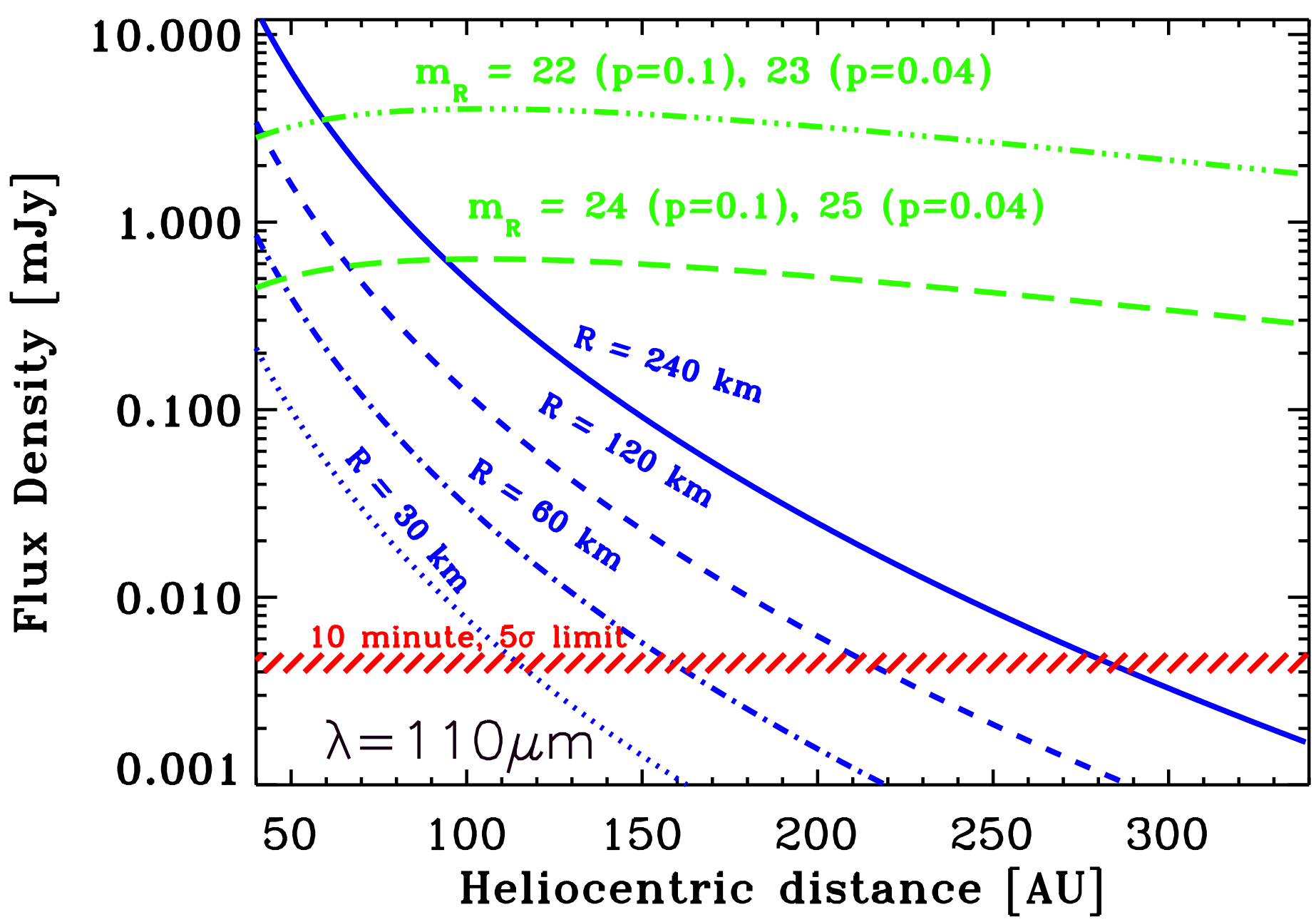

**Figure 1:** Flux density at 110 μm from TNOs of different radii (curves labeled by radius R) compared with CALISTO’s 5σ detection limits for integration times of 10 min (diagonal-dashed horizontal line) The curves with indicated R-band (620 nm) magnitude $m_R$, and geometric albedo, p, give the flux from TNOs which are at limit of optical detectability. These lie well above the CALISTO limits.

*A FIR TNO Survey*: Surveys of these more distant solar system bodies to date are limited by optical-band sensitivities down to the 22-24 magnitude level (cf. Petit et al. 2011) and sizes in excess of 100 km (cf. Vilenius et al. 2012). Even The Large Synoptic Survey Telescope (LSST) will have a limiting magnitude near this range ($m_R$~24.5; LSST Science Book V2. 2009, p. 18). A far-infrared (FIR) mission with survey capabilities, like the prospective Cryogenic Aperture Large Infrared Space Telescope Observatory (CALISTO; Goldsmith et al. 2008), offers the potential for the first time of really probing the population of TNOs down to moderates sizes, and out to distances exceeding 100 AU from the Sun.

Orbital Resonances with Neptune pump up inclination in the KBOs. Beyond 100AU, the TNO population may flare out as well, with a larger dispersion in inclination, and an increase in the surface density of objects (Morbidelli & Brown 2004). The green curves in Figure 1 (labeled with value of $m_R$) give an idea of the flux density produced by TNOs which are at the limit of detectability at optical wavelengths. CALISTO evidently can go more than one order of magnitude below this, even with predicted confusion limits, indicating the advantage of high sensitivity submillimeter observations of TNO thermal emission, and may go fainter with repeated observations of the field when the object has moved off of background sources.

The ability to derive large quantities of size measurements is a unique value of such FIR surveys. Small bodies typically can vary in their surface reflectivity by factors of 5 or more, while surveys that detect emitted light provide reliable sizes from the flux (cf. Mainzer et al. 2011). This is important because the previous optical surveys have provided alternate size frequency distributions, based on inferences of reflectivity, indicative of competing evolution histories for these bodies (Trujillo et al. 2001; Bernstein et al. 2004; Schlichting et al. 2013), especially at the smaller end (TNO diameters <100 km) of the size scales. Objects at TNO distances will be best detected at wavelengths near 110 $\mu$m. Shorter (~50 $\mu$m), and longer (~200 $\mu$m), wavelengths will better constrain the sizes and temperatures of the objects observed.

*Expected Populations:* Presently, most surveys have placed order-of-magnitude constraints on larger TNOs, with solar-system absolute magnitudes (H)~9 and sizes ~100 km. Petit et al. (2011) place the total of all TNOs, mostly in the classical KBO population near the ecliptic plane, over 100 km in size at ~130,000 in number, and Scattered Disk Objects (SDOs) down to similar sizes, more widely distributed in orbital inclination, near 25,000 in number. Schlichting et al. (2013) and Trujillo et al. (2001) place a cumulative size frequency distribution exponent value q~4, where the number of bodies N with diameters > D go as:

$$N(>D) \propto D^{1-q}$$

so that if, as Figure 1 suggests, a CALISTO-type survey of 1/10$^{th}$ of the sky is sensitive down to TNO diameters D~50 km or smaller, such a survey may yield several tens of thousands of new TNO discoveries, and a correspondingly large sample of TNO sizes, as well as thousands of new SDOs and diameters.

*Related Activity in Related Populations:* CALISTO also has the potential for detecting the limits of cometary activity in these and related populations. Species such as CO may drive sublimation out to distances of several tens of AU (Meech and Svoren, 2004). Detection of extended moving objects within a field owing to the presence of gas and dust coma is possible, and the expected size of such features would extend over several beam widths (A. J. Lovell, private communication). Such observations would place key constraints on the rates of mass lost to ejection of dust from these bodies, as well as the abundance of rarely-observed extremely-volatile species that may be relatively depleted in short-period comets (cf. Bauer et al. 2011, 2012). The onset of such distant activity may be linked to observational phenomenon heretofore unexplained, such as the source of the Centaur color bimodality (the red and gray sub-populations; Tegler et al. 2008), as well as place constraints on the primordial conditions under which they were formed.

**References:**

Barucci, M. A., H. Boehnhardt, D. P. Cruikshank, & A. Morbidelli 2008, in The Solar System Beyond Neptune: Overview and Perspective, ed. M. A. Barucci, H. Boehnhardt, D. P. Cruikshank, & A. Morbidelli (Tucson, AZ: Univ. Arizona Press), 3.

Bauer, J. M., Walker, R. G., Mainzer, A. K., et al. 2011. WISE/NEOWISE Observations of Comet 103P/Hartley 2. ApJ, 738, 171.

Bauer, J. M., Kramer, E. Mainzer, A. K., et al. 2012. WISE/NEOWISE Preliminary Analysis and Highlights of the 67P/Churyumov-Gerasimenko near Nucleus Environs. ApJ, 758, 18.

Bernstein, G. M. D. E. Trilling, R. L Allen, et al. 2004. The Size Distribution of Trans-Neptunian Bodies, AJ 128, 1364.

Gladman, B., B. G. Marsden, and C. VanLaerhoven 2008, in The Solar System Beyond Neptune, Nomenclature in the Outer Solar System, ed. M. A. Barucci, H. Boehnhardt, D. P. Cruikshank, & A. Morbidelli (Tucson, AZ: Univ. Arizona Press), 43.

Goldsmith, P. F., M. Bradford, M. Dragovan, C. Paine, C. Satter, et al. "CALISTO: the Cryogenic Aperture Large Infrared Space Telescope Observatory", Proc. SPIE 7010, Space Telescopes and Instrumentation 2008: Optical, Infrared, and Millimeter, 701020 (July 12, 2008); doi:10.1117/12.788412

Mainzer et al. 2011. NEOWISE Observations of Near-Earth Objects: Preliminary Results. ApJ. 743, 156.

Meech, K. J., and J. Svoren. 2004. in Comets II, Using Cometary Activity toTrace the Physical and Chemical Evolution of Cometary Nuclei, ed. M. C.Festou et al. (Tucson, AZ: Univ. Arizona), 317.

Morbidelli, A. and M. E. Brown. 2004. in Comets II, The kuiper belt and the primordial evolution of the solar system, ed. M. C.Festou et al. (Tucson, AZ: Univ. Arizona), 175.

Petit, J.-M., J. J. Kavelaars, B. J. Gladman, R. L. Jones, J. Wm. Parker, C. Van Laerhoven, P. Nicholson, G. Mars, P. Rousselot, O. Mousis, B. Marsden, A. Bieryla, M. Taylor, M. L. N. Ashby, P. Benavidez, A. Campo Bagatin, and G. Bernabeu. 2011. The Canada France Ecliptic Survey – Full Data Release: The Orbital Structure of the Kuiper Belt. Astron. J. 142, 131.

Schlichting, H. E., C. I Fuentes, and D. E. Trilling. 2013. Initial Planetesimal Sizes and the Size Distribution of Small Kuiper Belt Objects. AJ 146, 36.

Tegler, S. C., J. M. Bauer, W. Romanishin, and N. Peixinho. 2008, in The Solar System Beyond Neptune, Colors of Centaurs, ed. M. A. Barucci, H. Boehnhardt, D. P. Cruikshank, & A. Morbidelli (Tucson, AZ: Univ. Arizona Press), 105.

Trujillo, C. A., D. C. Jewitt, and J. X. Luu, 2001. Properties of the Trans-Neptunian Belt: Statistics from the Canada-France-Hawaii Telescope Survey. AJ 122, 457.

Vilenius, E., C. Kiss, M. Mommert, T. Müller, P. Santos-Sanz, et al. 2012. “TNOs are Cool”: A survey of the trans-Neptuneian region, VI. Herschel⋆/PACS observations and thermal modeling of 19 classical Kuiper belt objects. Astron. & Astroph. 541, A94.